%% file: LHY1D.tex
\documentclass[aps,prl,twocolumn]{revtex4-1}
\usepackage[pdfview=FitV,pdfstartview=FitV]{hyperref}
\usepackage{amsmath,amssymb,amsfonts}
\usepackage{lmodern}
\usepackage{empheq} 
\usepackage{bm} 
\usepackage{graphicx}
\usepackage{color}
\usepackage{tikz}
\usepackage{csquotes}

\renewcommand{\d}[2][]{\mathop{\textnormal{d}^{#1} #2}}
\newcommand{\im}{\mathrm{i}}

\begin{document}
\title{Quantum fluctuations in quasi-one-dimensional dipolar Bose-Einstein condensates}
\author{D. Edler\textsuperscript{1}, C. Mishra\textsuperscript{1,2}, F. W\"achtler\textsuperscript{1},  R. Nath\textsuperscript{2}, S. Sinha\textsuperscript{3}, and L. Santos\textsuperscript{1}}
\affiliation{
\textsuperscript{1}Institut f\"ur Theoretische Physik, Leibniz Universit\"at, 30167 Hannover, Germany,
\textsuperscript{2}Indian Institute of Science Education and Research, Pune 411 008, India,
\textsuperscript{3}Indian Institute of Science Education and Research-Kolkata, Mohanpur, Nadia-741246, India}
\date{\today}

\begin{abstract}
Recent experiments have revealed that beyond-mean-field corrections are much more relevant in weakly-interacting dipolar condensates than in their non-dipolar counterparts. We show that in quasi-one-dimensional geometries quantum corrections in dipolar and non-dipolar condensates are strikingly different due to the peculiar momentum dependence of the dipolar interactions. The energy correction of the condensate presents not only a modified density dependence, but it may even change from attractive to repulsive at a critical density due to the surprising role played by the transversal directions. The anomalous quantum correction translates into a strongly modified physics for quantum-stabilized droplets and dipolar solitons. Moreover, and for similar reasons, quantum corrections of three-body correlations, and hence of three-body losses, are strongly modified by the dipolar interactions. This intriguing physics can be readily probed in current experiments with magnetic atoms.
\end{abstract}

\maketitle

\doi{10.1103/PhysRevLett.119.050403}



\paragraph{Introduction.--}  


Quantum fluctuations introduce a shift of the ground-state energy of a Bose gas, which at first order is given by the well-known Lee-Huang-Yang~(LHY) correction~\cite{Lee1957}. However, in the weakly-interacting regime, experiments on Bose-Einstein condensates are well described within the mean-field approximation.  The situation may be crucially different in the presence of competing interactions, as recently discussed in the context of Bose-Bose mixtures~\cite{Petrov2015}. In that scenario, the interplay between inter- and intra-species interactions results, at the verge of mean-field instability, in a dominant LHY correction well within the weakly-interacting regime.  The LHY correction may stabilize a collapsing condensate, resulting in the formation of quantum droplets, a novel ultra-dilute liquid whose surface tension is provided by purely quantum effects.  


Dipolar condensates, formed by particles with large magnetic or electric dipolar moments, are also characterized by competing interactions, in this case short-range and dipole-dipole interactions. Indeed, recent experiments on highly magnetic atoms have revealed the crucial role played by quantum fluctuations at the mean-field instability, showing for the first time the formation of quantum droplets~\cite{Kadau2016}, which may remain self-bound even in the absence of external trapping~\cite{Schmitt2016}.  Quantum stabilization and droplet formation have attracted wide theoretical and experimental attention~\cite{Ferrier2016,Waechtler2016,Bisset2016, Saito2016,Waechtler2016b,Boudjemaa2016, Baillie2016, Chomaz2016}, being a general phenomenon that is expected to characterize not only condensates of magnetic atoms, but the whole rapidly developing field of strongly dipolar gases~\cite{Lahaye2009,Baranov2012}.


In Bose-Bose mixtures and in dipolar condensates quantum stabilization stems from the compensation between the attractive residual mean-field interaction, proportional to the 3D density $n_\text{3D}  $,  and the repulsive LHY correction, which in both systems is proportional to $n_\text{3D}  ^{3/2}$~\cite{Petrov2015,Lima2011}.  As a result, there is a critical density at which both contributions compensate.  Quantum fluctuations play an even more intriguing role in lower dimensions. In particular, droplets are stabilized for a sufficiently low density in 1D Bose-Bose mixtures~\cite{Petrov2016}, against melting rather than collapse, by the competition of a residual repulsive mean-field term, proportional to the 1D density $n_\text{1D}$, and the attractive LHY correction, proportional to $-n_\text{1D}^{1/2}$.

Whereas beyond mean-field effects in 3D Bose-Bose mixtures and dipolar condensates are very similar due to the almost identical density dependence of the quantum correction, we show in this Letter that quantum fluctuations lead in quasi-1D dipolar condensates to a strikingly different physics compared to their non-dipolar counterparts.  This difference stems from the peculiar momentum dependence of the dipole-dipole interactions in quasi-1D geometries~\cite{Sinha2007}.  As a result, not only is the density dependence of the quantum corrections very different, but even its sign may change due to the remarkable role played by transversal directions in dipolar gases well within the 1D regime.  The anomalous quantum corrections change the nature of quantum stabilization and strongly influence the physics of solitons.  We also show that, whereas three-body correlations present the same density dependence in 3D dipolar and non-dipolar condensates~\cite{Kagan1985}, they display in 1D a radically different dependence.


\paragraph{Dipolar interaction in 1D.--} We consider bosons with mass $M$ and magnetic moment $\vec\mu_D$, although our results also apply for electric dipoles.  The system is strongly confined on the $xy$ plane by an isotropic harmonic trap of frequency $\omega_\perp$, but it is untrapped along $z$.  We assume that the chemical potential $|\mu| \ll \hbar\omega_\perp$, and hence the condensate remains kinematically 1D such that its wave function splits as $\Psi(\vec r)=\psi(x,y)\phi(z)$, with $\psi(x,y)=e^{-(x^2+y^2)/2l_\perp^2}/{\sqrt{\pi}l_\perp}$ the ground state of the transversal trap, with $\l_\perp^2=\hbar/{M\omega_\perp}$.  After integrating over $x$ and $y$ the interaction between particles in the condensate acquires a momentum, $k$, dependence of the form
\begin{align}
  \tilde V_\text{1D}(k) = g_\text{1D} \left\{
    1 + \varepsilon_\text{dd}   \left [3 F \left (0,k^2l_\perp^2/2 \right ) -1 \right ]
  \right\},
\end{align}
with $F(j,\sigma)\equiv \sigma^{j+1} e^\sigma \Gamma(-j,\sigma)$~\cite{Sinha2007}, where $\Gamma(-j,\sigma)$ is the incomplete Gamma function. Short-range interactions are characterized by the 1D coupling constant $g_\text{1D}  = g_\text{3D}  /2\pi l_\perp^2$, where $g_\text{3D}  =4\pi\hbar^2 a /M$, with $a>0$ the $s$-wave scattering length.  Assuming $\vec\mu_D$ along $z$, $\varepsilon_\text{dd}  =\mu_0\mu_D^2 /{3 g_\text{3D}}  $ is the ratio between the strengths of the dipolar and contact interactions~\cite{footnote-alpha}, with $\mu_0$ the vacuum permeability. This 1D condition $|\mu|/{\hbar\omega_\perp} \ll 1$ demands $|1-\varepsilon_\text{dd}|\ll 1/2 n_\text{1D} a$, a condition satisfied in all the calculations in this paper~\cite{footnote-1D}.

\begin{figure}[t]
\begin{center}
  \begin{tikzpicture}
    \definecolor{lightblue}{RGB}{96, 149, 201}
    \draw[color=lightblue, ultra thick, scale=0.5, <-]
         (2,1) -- (0,0);
    \draw[color=lightblue, ultra thick]
         (2,1) node[left, black, xshift=1.25em, yshift=0.75em] {$(n'_r, m, k)$} -- (0,0);

    \draw[color=lightblue, ultra thick, scale=0.5, >-]
         (-2,1) -- (0,0);
    \draw[color=lightblue, ultra thick]
         (-2,1) node[right, black, xshift=-1.25em, yshift=0.75em] {$(n_r, m, k)$} -- (0,0);
    \shade[ball color=red] (0,0) circle[radius=0.25cm];

    \begin{scope}[xshift=3.5cm]
      \draw[color=lightblue, ultra thick, scale=0.5, <-]
           (2,1) -- (0,0);
      \draw[color=lightblue, ultra thick]
           (2,1) node[left, black, xshift=-1em] {$(n_r, m, k)$} -- (0,0);

      \draw[color=lightblue, ultra thick, scale=0.5, <-]
           (2,-1) -- (0,0);
      \draw[color=lightblue, ultra thick]
           (2,-1) node[left, black, xshift=-1em] {$(n_r, -m, -k)$} -- (0,0);
      \shade[ball color=red] (0,0) circle[radius=0.25cm];
    \end{scope}
  \end{tikzpicture}
\end{center}
\vspace{-0.4cm}
\caption{Dominant collisions between particles in the condensate and in excited states~(see text).}
\label{fig:2n}
\end{figure}
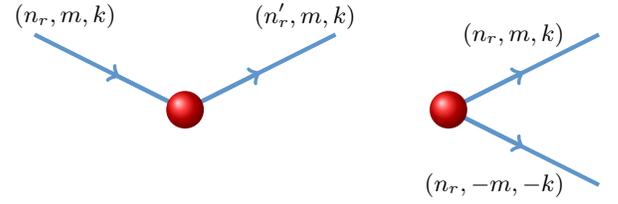

\paragraph{LHY correction.--}  Single particle excitations, $(n_r,m,k)$, are characterized by their radial quantum number $n_r$, angular momentum $m$, and axial linear momentum $k$.  In 1D contact-interacting systems transversal excitations, with $(n_r,m)\neq (0,0)$, play a negligible role in beyond-mean-field corrections.  This may be crucially different in dipolar gases.  In the weakly-interacting regime, the main processes involving condensed and excited particles are sketched in Fig.~\ref{fig:2n}.  A collision between a particle in $(n_r,m,k)$ and one in the condensate $(0,0,0)$ (left), preserves both $m$ and $k$, but may change the radial number into $n'_r$.  On the other hand two condensed particles may collide (right) and create excitations in $(n_r,m,k)$, and $(n'_r,-m,-k)$.  Both processes are characterized by the interaction energy~\cite{footnote-SM}
\begin{align}
 \left(\hat U_{m}(k)\right)_{n_r,n'_r} \!\!= g_\text{1D} n_\text{1D}   C_{n_r, n'_r, m} F\left (n_r\!+\!n'_r\!+\!m,\frac{k^2l_\perp^2}{2} \right ),
\end{align}
where $C_{n_r, n'_r, m}=6 \frac{(-1)^{n_r+n'_r}}{2^{n_r+n'_r+m+1}} \sqrt{ \binom{n_r+n'_r+m}{n_r} \binom{n_r+n'_r+m}{n'_r} }$, and we have considered for simplicity 
$\varepsilon_\text{dd}  =1$~\cite{footnote-comment-edd}. It is crucial that, although for $\varepsilon_\text{dd}  =1$ the compensation of dipolar and contact interactions results in an ideal 1D condensate~($\tilde V_\text{1D}(0) = \big(\hat U_{0}(0)\big)_{0,0}=0$), $\big(\hat U_{m}(k)\big)_{n_r,n'_r}$ may be of the order of $g_\text{1D} n_\text{1D}  $. Because of this peculiar feature, which stems from the momentum dependence of the dipolar interactions, transversal excitations play in dipolar gases a key role in quantum corrections if $g_\text{1D}   n_\text{1D}   \gtrsim \hbar \omega_\perp$ despite the 1D character of the condensate. 



\begin{figure}[t]
  \begin{center}
    \def\svgwidth{0.9\columnwidth}
    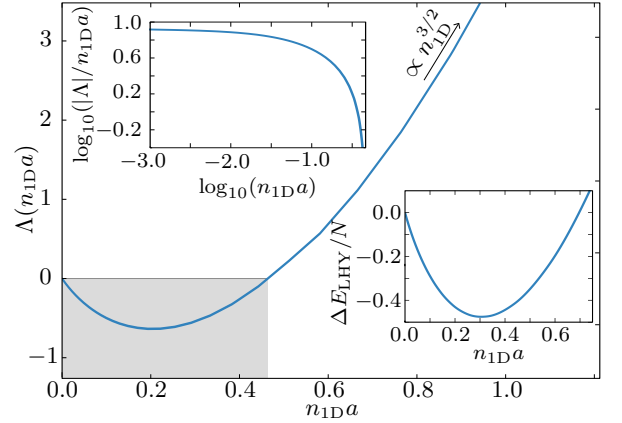
  \end{center}
  \vspace{-0.4cm}
  \caption{
    $\Lambda \equiv  \frac{\Delta\mu_\text{LHY}  }{\hbar\omega_\perp}\frac{l_\perp}{a}$ as a function of $n_\text{1D} a$ for an homogeneous 1D dipolar condensate. The top inset depicts $\log_{10}(|\Lambda|/{n_\text{1D} a})$ against $\log_{10}(n_\text{1D} a)$, showing that $\Delta\mu_\text{LHY}  \propto n_\text{1D}  $ for $n_\text{1D}  \to 0$. The bottom inset shows $\Delta E_\text{LHY}/N$ as a function of $n_\text{1D}a$.}
  \label{fig:3n}
\end{figure}


The elementary excitations may be obtained for each $m$ and $k$ from the Bogoliubov-de Gennes equations:
\begin{align} 
  \xi_\nu \begin{pmatrix}
    \vec u_\nu  \\
    \vec v_\nu
  \end{pmatrix} = \begin{pmatrix}
    \hat E_m(k) \!+\! \hat U_m(k) & \hat U_m(k)  \\
    -\hat U_m(k) & \!\! -\hat E_m(k)\! -\! \hat U_m(k)
  \end{pmatrix}
  \begin{pmatrix}
    \vec u_\nu  \\
    \vec v_\nu
  \end{pmatrix},
\end{align}
where $\big(\hat E_m(k)\big)_{n_r,n'_r}=E_{n_r m}(k)\delta_{n_r, n'_r}$, with $E_{n_r m}(k)=\hbar k^2/{2M} + \hbar\omega_\perp (2n_r+m)$.
Following a similar procedure as in Ref.~\cite{Hugenholtz1959} the LHY energy correction, $\Delta E_\text{LHY}  $, 
may be obtained from the differential equation~\cite{footnote-SM}: 
\begin{align}
  &\frac{\Delta E_\text{LHY}  }{L} -\frac{1}{2} n_\text{1D}   \frac{\d{}}{\d{n_\text{1D}}  } \left (\frac{\Delta E_\text{LHY}  }{L}\right ) \nonumber \\
 =&\frac{1}{2}\sum_m \int_{-\infty}^{\infty} \frac{\d{k}}{2\pi} \sum_\nu \sum_{n_r} \left[ E_{n_r  m} (k)- \xi_\nu \right] (\vec v_\nu)_{n_r}^2,
\label{eq:HP}
\end{align}
with $L$ the quantization length~\cite{footnote-weaklyinteracting}.  Figure~\ref{fig:3n} shows the LHY correction of the chemical 
potential, $\Delta \mu _\text{LHY}  =\frac{\d{}}{\d{n_\text{1D}}} \left (\frac{\Delta E_\text{LHY}  }{L}\right )$, for different $g_\text{1D}   n_\text{1D}   / 2\hbar\omega_\perp = n_\text{1D}   a$. 

For $n_\text{1D}  a \ll 1$, the effect of the transversal modes is, as expected, negligible, and the LHY correction remains attractive. However, whereas for contact interacting systems $\Delta\mu_\text{LHY}  \propto -n_\text{1D}^{1/2}$~\cite{Petrov2016}, the density dependence in dipolar condensates is radically different.  For $n_\text{1D}  a\to 0$, $\Delta\mu_\text{LHY}   \propto -n_\text{1D}$, whereas for growing $n_\text{1D}a$, $\Delta \mu_\text{LHY}$ departs from the linear dependence~(top inset of Fig.~\ref{fig:3n}). This is crucial for the physics of 1D droplets, as discussed below. 

For  $n_\text{1D}  a \gtrsim 0.1$, transversal excitations become significant.  The LHY correction reaches a maximal negative value at   $n_\text{1D}  a\simeq 0.2$, and then increases, becoming repulsive for $n_\text{1D}  a>0.42$.  For $(n_\text{1D}  a)\gg 1$, $\Delta\mu_\text{LHY}   \propto n_\text{1D}  ^{3/2}$, i.e.\ the LHY correction becomes that expected for a 3D condensate~\cite{Lima2011}.  This radical change in the nature of the quantum correction for a condensate well within the 1D regime constitutes a striking qualitatively novel feature of quasi-1D dipolar gases.

\paragraph{Phase diagram.-- }
We consider at this point an axially un-trapped but possibly self-bound condensate, with an axial width $R\gg l_\perp$.  In that case the use of the local density approximation, i.e.\ substituting in Eq.~\eqref{eq:HP} $n_\text{1D}$ by $n_\text{1D}  (z)$, is well justified since the momenta contributing most to the LHY correction fulfill $kR\gg 1$.
The resulting generalized Gross-Pitaevskii equation is
\begin{align}
  \mu\phi(z)=&\frac{-\hbar^2}{2M}\frac{\d[2]{\phi}}{\d[2]{z}}\nonumber\\
            &+ \phi(z) \big\{\mu_\text{MF}  [n_\text{1D}  (z)]+\Delta\mu_\text{LHY}  [n_\text{1D}  (z)] \big\}
\label{eq:GPE}
\end{align}
with $\mu$ the chemical potential, and $\mu_\text{MF}  [n_\text{1D}(z)]=\int \frac{\d{k}}{2\pi}\tilde V_\text{1D}  (k)\tilde n_\text{1D}  (k)e^{\im kz}$ the mean-field interaction, with $\tilde n_\text{1D}  (k)$ the Fourier transform of $n_\text{1D}  (z)$.

\begin{figure}[t]
  \begin{center}
    \def\svgwidth{0.90\columnwidth}
    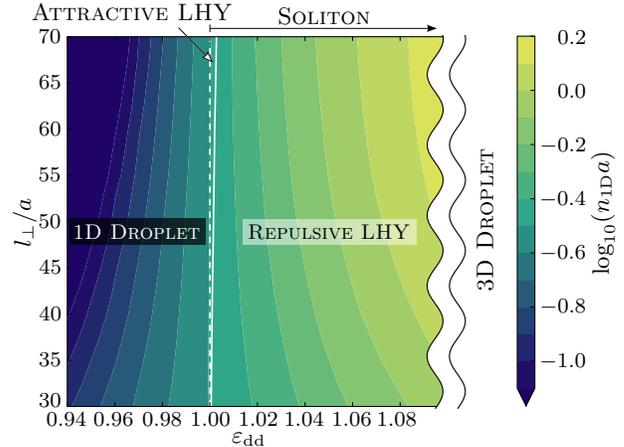
  \end{center}
  \vspace{-0.4cm}
  \caption{Peak density for $N=5000$ atoms as a function of $l_\perp/a$ and $\varepsilon_\text{dd}$.  At $\varepsilon_\text{dd}  >1$~(dashed vertical line) the droplet regime smoothly crossovers into the soliton regime.  When $n_\text{1D}  a\ll 1$, the anomalous density dependence of $\Delta\mu_\text{LHY}$ results in droplet melting. The solid line marks the point at which the LHY becomes in average repulsive.}
  \label{fig:0}
\end{figure}

\begin{figure}[t]
  \begin{center}
    \includegraphics[width=0.9\columnwidth]{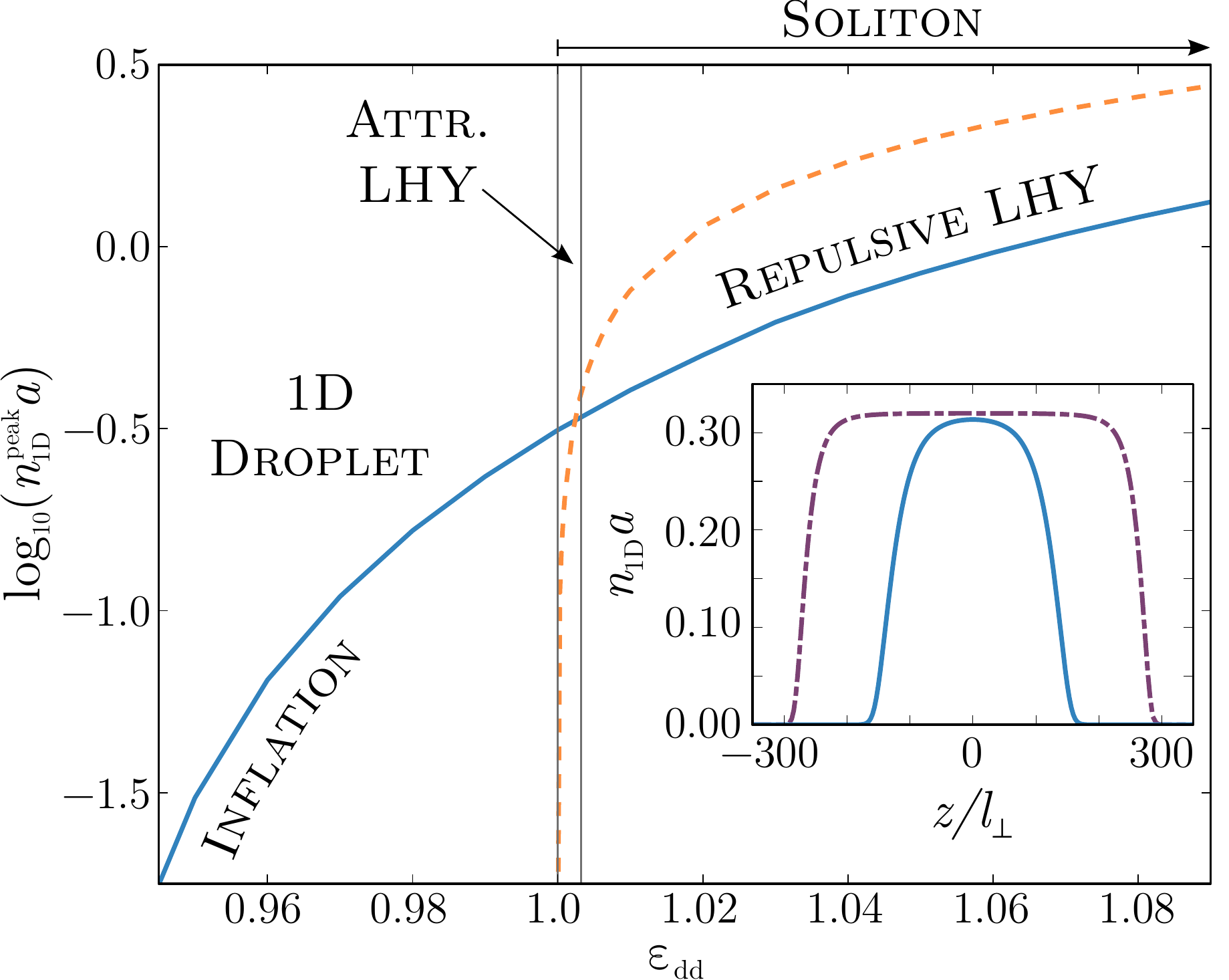}
  \end{center}
  \vspace{-0.4cm}
  \caption{Peak density,  $n_\text{1D}^{\mathrm{peak}}$, of a 1D self-bound dipolar condensate of $N=5000$ particles with $l_\perp/a=65$ as a function of $\varepsilon_\text{dd}  $. The dashed and solid curves depict, respectively, the mean-field results, and those taking into account the LHY correction. For a discussion of the different regions, see text. In the inset we depict a typical flat-top droplet profile for $\varepsilon_\text{dd}  =1$ with $l_\perp/a=65$~(solid) and $30$~(dot-dashed).}
  \label{fig:4}
\end{figure}

Figure~\ref{fig:0} depicts the peak density for $N=\int_{-\infty}^\infty \d{z} n_\text{1D}(z)=5000$ particles as a function of $\varepsilon_\text{dd}$ and $l_\perp/a$~(which must be $\gg 1$ to guarantee the 3D nature of the scattering~\cite{footnote-Olshanii}). Neglecting quantum corrections, the interactions are repulsive for $\varepsilon_\text{dd}  < 1$ preventing any self-bound solution~(see Fig.~\ref{fig:4}), whereas for  $\varepsilon_\text{dd}  >1$ the attractive interactions lead to the formation of a soliton.

For sufficiently low densities, the effective LHY attraction results as in Bose-Bose mixtures~\cite{Petrov2016} in the formation for $\varepsilon_\text{dd}  \leq 1$ of self-bound droplets~(see Figs.~\ref{fig:0} and~\ref{fig:4}) that  present a flat top profile~(inset of Fig.~\ref{fig:4}).  Note that at $\varepsilon_\text{dd}  =1$, the mean-field contribution vanishes. The droplet acquires, however, a finite peak density, $n_\text{1D}^{\text{peak}} a \simeq 0.3$~(inset of Fig.~\ref{fig:4}), at which $\Delta E_\text{LHY}  /N$ is minimal~(bottom inset of Fig.~\ref{fig:3n}).  Note that this minimum, and hence the universal peak droplet density at $\varepsilon_\text{dd}=1$, also results from the nontrivial role played by the transversal degrees of freedom.

When $\varepsilon_\text{dd}  $ is lowered, the density decreases and the system enters in the regime in which $\Delta\mu_\text{LHY}  \propto -n_\text{1D} $.  Since the LHY correction and the mean-field energy have an equal density dependence, the competition between both energies crucial for quantum stabilization is absent, and the system undergoes an abrupt \emph{droplet inflation} into the unbound solution.  The latter must be compared to the case of non-dipolar Bose-Bose mixtures, which are characterized by a fixed dependence $\Delta\mu_\text{LHY}  \propto -n_\text{1D}^{1/2}$.  As a result, the competition between mean-field energy and LHY correction remains efficient in binary mixtures even at very low densities, and hence the peak density smoothly decreases within the mean-field unbound region without any droplet inflation.



At $\varepsilon_\text{dd} = 1$ the system smoothly crossovers into the soliton regime. For $\varepsilon_\text{dd} > 1$ the soliton density grows smoothly for increasing $\varepsilon_\text{dd} $, and the LHY correction changes eventually from attractive to repulsive. When this occurs the soliton density is significantly lower than that expected from mean-field theory~(up to a factor of $2$ in Fig.~\ref{fig:4}). Moreover, since for large-enough densities, $\Delta\mu_\text{LHY}  \propto n_\text{1D}^{3/2}$, the effect of the LHY correction remains relevant even far from the mean-field instability. This must be compared to the case of Bose-Bose mixtures, where the $\Delta\mu_\text{LHY}  \propto -n_\text{1D}  ^{1/2}$ dependence renders the LHY correction basically negligible within the soliton regime. Note that for sufficiently large $\varepsilon_\text{dd}  >1$, eventually $\mu\gtrsim \hbar\omega_\perp$, and the condensate crossovers into the 3D regime, where the repulsive LHY prevents collapse.  This would correspond to the elongated 3D macro-droplet regime recently explored experimentally~\cite{Chomaz2016}.  The description of this crossover lies, however, beyond the scope of this paper.


\paragraph{Three-body correlations.--} 
Whereas in mean-field approximation three-body correlations fulfill $g^{(3)}=\frac{1}{n(\vec r)^3}\langle \hat\Psi^\dag (\vec r)^3\hat\Psi(\vec r)^3 \rangle = 1$, quantum corrections may significantly correct its value, $g^{(3)} = 1 + \Delta g^{(3)}$, and hence in turn the three-body loss rate. For homogeneous 3D non-dipolar condensates with density $n_\text{3D}  $,  $\Delta g^{(3)}\simeq \frac{64}{\sqrt{\pi}}(n_\text{3D}   a^3)^{1/2}$~\cite{Kagan1985}, as confirmed in recent experiments~\cite{Haller2011}.  As for the LHY correction, in 3D homogenous dipolar condensates,  the correction of $g^{(3)}$ is very similar:  
$\Delta g^{(3)}\simeq \frac{64}{\sqrt{\pi}}(n_\text{3D}  a^3)^{1/2} (1+C \varepsilon_\text{dd}  ^2)$, with $C\simeq 0.3$ \cite{footnote-g3-3D}. 
Dipolar interactions hence introduce corrections that may be sizable in current experiments with magnetic atoms, but the density-dependence of $g^{(3)}$ is identical to that of non-dipolar condensates. 

\begin{figure}[t]
\begin{center}
\includegraphics[width=0.9\columnwidth]{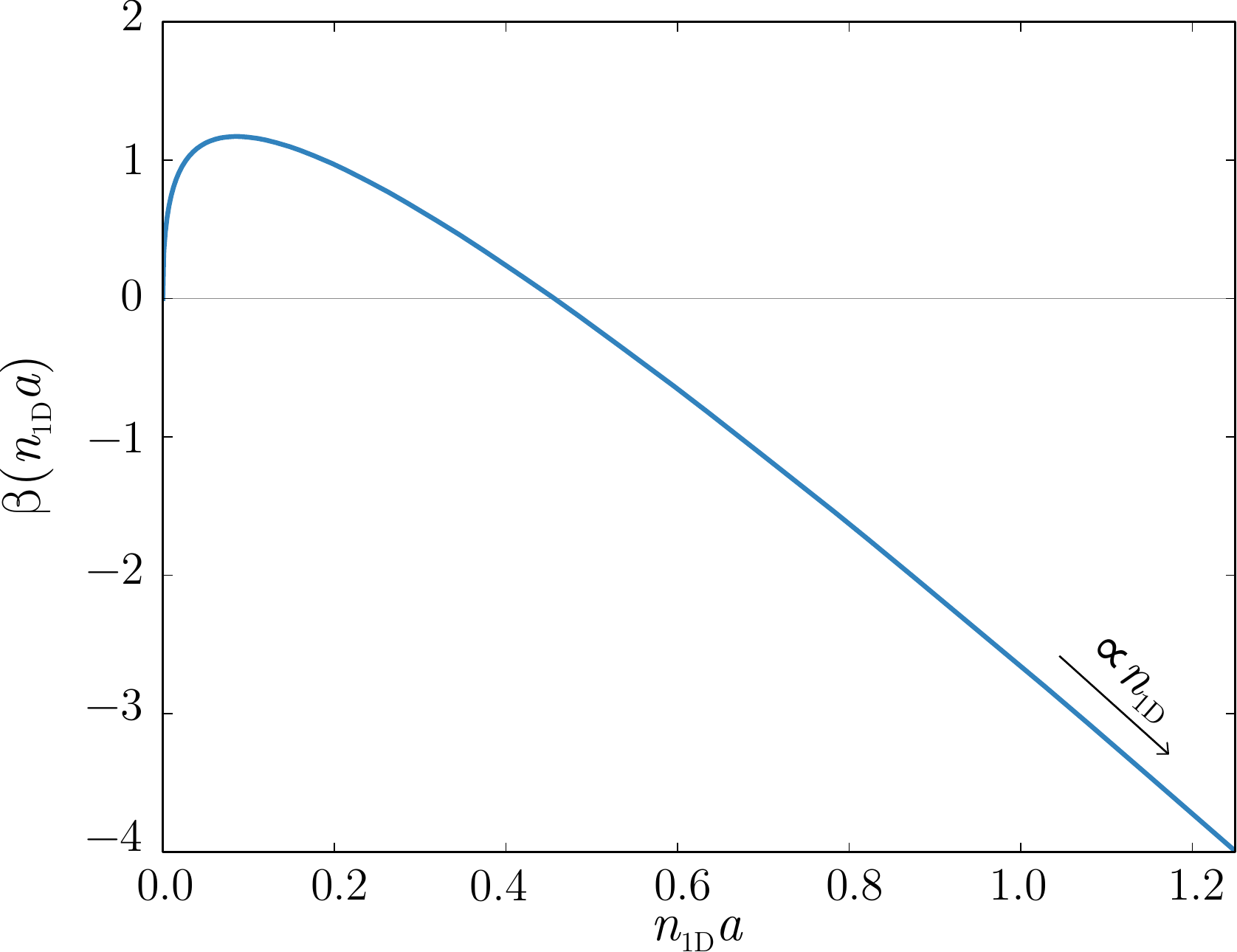}
\end{center}
\vspace{-0.4cm}
\caption{(Color online) Correction of the three-body correlations, $\beta(\varepsilon_\text{dd}  =1,n_\text{1D}  a)=-\Delta g^{(3)}\frac{\pi}{6\sqrt{\gamma}}$.}
\label{fig:5}
\end{figure}

The situation is radically different in 1D. 
For a 1D non-dipolar condensate $\Delta g^{(3)}\!\! =\!-\frac{6}{\pi}\sqrt{\gamma}$ \cite{Gangardt2003}, with $\gamma=2a/{n_\text{1D}  l_\perp^2}\!\ll\! 1$.  Three-body correlations are hence reduced by quantum effects, and the correction increases for a decreasing density, since, counter-intuitively, 1D systems are more strongly interacting the more dilute they are.  As for the LHY correction, the momentum dependence of the dipolar interactions leads to a markedly different density dependence in dipolar condensates.  The correction of $g^{(3)}$ averaged over the transversal degree of freedom~\cite{footnote-g3} may be evaluated from the LHY correction using the Hellmann-Feynman theorem~\cite{footnote-SM}:
\begin{align}
  \Delta g^{(3)} \equiv& \int \frac{\d[3]r}{L} \frac{\psi(x,y)^4}{\int \d{x'} \d{y'} \psi(x',y')^4} \left ( 
\frac{\langle \hat\Psi^\dag (\vec r)^3\hat\Psi(\vec r)^3 \rangle}{n(\vec r)^3} -1 \right ) \nonumber \\
  =& \frac{6}{n_\text{1D}  ^2 L}\frac{\partial\Delta E_\text{LHY}  }{\partial g_\text{1D}  } = -\frac{6}{\pi}\sqrt{\gamma} \beta (\varepsilon_\text{dd}, n_\text{1D}  a),
\end{align}
where $ \beta (\varepsilon_\text{dd}  ,n_\text{1D}  a)$ is depicted in Fig.~\ref{fig:5}. 
For small $n_\text{1D}  a$, $\Delta g^{(3)} \propto - n_\text{1D}  ^\lambda$, with $-1/2<\lambda<0$. As for non-dipolar condensates $\Delta g^{(3)}$ remains negative and increases with decreasing $n_\text{1D}  $, albeit with a significantly modified power law. In contrast, when $n_\text{1D}   a>0.42$, the growing role of the transversal modes results into a change in the sign of $\Delta g^{(3)}$, i.e.\ three-body correlations are enhanced rather than reduced by quantum effects despite of the fact that the condensate remains in the 1D regime. For $n_\text{1D}   a \gg 1$, $\Delta g^{(3)} \propto n_\text{1D}  ^{1/2}$,  as expected for 3D condensates.  This non-trivial behavior of three-body correlations in quasi-1D dipolar condensates may be probed in on-going experiments with magnetic atoms using similar techniques as those applied in non-dipolar quasi-1D condensates~\cite{Haller2011}.



\paragraph{Conclusions.--} 
The momentum-dependence of the dipolar interactions leads to strikingly different quantum effects in quasi-one-dimensional dipolar condensates compared to their non-dipolar counterparts. In contrast to Bose-Bose mixtures, quantum stabilization is disrupted in dipolar condensates at low densities due to the modified density dependence of the LHY correction.  As a result quantum droplets only exist in a window of density values. Moreover, although the condensate remains one-dimensional, the LHY may be crucially affected by transversal modes, which induce a change from attractive to repulsive LHY correction at a critical density.  This change of character results in a significant reduction of the peak density of the soliton, as well as a modification of its shape. Hence quantum corrections should be carefully considered in future studies of dipolar solitons. Furthermore, the peculiar nature of quantum fluctuations is also reflected in the beyond-mean-field correction of three-body losses, which also changes its sign within the 1D regime for growing density.  Our results open intriguing questions about 2D dipolar condensates, where we expect a similar non-trivial density dependence of the quantum corrections, as well as about the role of transverse modes in anharmonic transversal confinements. This surprising physics of low-dimensional dipolar condensates 
can be readily probed in current experiments  with magnetic atoms. 



We thank F.\ Ferlaino, L.\ Chomaz, S.\ Baier, I.\ Ferrier-Barbut, and T.\ Pfau for their insightful comments, and very especially, H.-P.\ B\"uchler for pointing out an error in a previous version. We acknowledge support by the DFG (RTG 1729) and DFG/FWF (FOR 2247), the Indo-French Centre for the Promotion of Advanced Research, the IP@Leibniz programme of the Leibniz Universit\"at Hannover, and the INSPIRE Fellowship Programme~(DST, India). C.~M.\ and S.~S.\ would like to thank the ITP Hannover for its hospitality.

The underlying raw data of all plots can be found on Zenodo.\cite{footnote-raw}

\end{document}


\title{Supplementary material for \\ ``Quantum fluctuations in one-dimensional dipolar Bose-Einstein condensates''}
\author{D. Edler$^1$, C. Mishra$^{1,2}$, F. W\"achtler$^1$,  R. Nath$^2$, S. Sinha$^3$, and L. Santos$^1$}
\affiliation{
$^1$ Institut f\"ur Theoretische Physik, Leibniz Universit\"at, 30167 Hannover, Germany,
$^2$ Indian Institute of Science Education and Research, Pune 411 008, India,
$^3$ Indian Institute of Science Education and Research-Kolkata, Mohanpur, Nadia-741246, India}

\maketitle


\section{Derivation of equation (2)} 

At this point we briefly comment on the calculation of the interaction potentials $U_{n_r,n'_r,m}(k)$. 
The interaction energy is of the form:
\begin{equation}
H_I=\frac{1}{2}\int d^3 r \int d^3 r' V(r-r') \hat \Psi(\vec r)^\dag \hat \Psi(\vec {r'})^\dag \hat \Psi(\vec {r'}) \hat \Psi(\vec r),
\end{equation}
where $\hat \psi(\vec r)$ the bosonic annihilation operator for a particle at the position $\vec r$, and $V(\vec r)$ is the 
interaction potential, containing both the short-range and dipole-dipole interactions. We may then express $\hat\Psi$ 
in terms of the modes $(n_r,m,k)$: $\hat \Psi(\vec r)=\sum_k \sum_{n_r,m} R_{n_r m}(\rho) e^{im\phi}e^{ikz} \hat a_{n_r, m,k}$, 
where the radial wave functions are
\begin{equation}
R_{n_r,m}(\rho)=\frac{(-1)^n}{l_\perp \sqrt{\pi}}\sqrt{\frac{n!}{(n+m)!}} \left (\frac{\rho}{l_\perp}\right )^m L_n^m \left (\frac{\rho^2}{l_\perp^2}\right ) e^{- \frac{\rho^2}{2l_\perp^2}},
\end{equation}
with $L_n^m$ the Laguerre polynomials. Assuming a condensate in $(n_r,m,k)=(0,0,0)$, the second-order contribution to the interaction Hamiltonian acquires the form:

\begin{widetext}
\begin{equation}
\frac{H_I^{(2)}}{L}=\frac{1}{2}\int \frac{dk}{2\pi}\sum_m \sum_{n,n'} U_{n_r,n'_r,m}(k) \left ( 2  \hat a_{n_r,m,k}^\dag  \hat a_{n'_r,m,k}+ 
 \hat a_{n_r,m,k}^\dag  \hat a_{n_r,-m,-k}^\dag +  \hat a_{n_r,m,k} \hat a_{n_r,-m,-k} \right ),
\end{equation}
\end{widetext}
where 
\begin{equation}
\!\!\!\frac{U_{n_r,n'_r,m}(k)}{g_\text{1D}  n_\text{1D}   (2\pi l_\perp)^2} \!=\! \! \int \! k_\rho d k_\rho \tilde V(k_\rho,k)\lambda_{n_r,m}(k_\rho)\lambda_{n'_r,m}(k_\rho),
\label{eq:U-1}
\end{equation}
where $g_\text{1D}  =g_\text{3D}   / 2\pi l_\perp^2$, $ \tilde V(k_\rho,k)=1+\epsilon_\text{dd}   \left ( 3\frac{k^2}{k^2 + k_\rho^2}-1 \right )$, 
and we introduce the $m$-th order Hankel transform of $R_{n_r,m}(\rho)R_{0,0}(\rho)$:
\begin{eqnarray}
\lambda_{n_r,m}(k_\rho)&=&\int d\rho \rho R_{n_r,m}(\rho)R_{0,0}(\rho)J_m(k_\rho \rho) \nonumber \\
&=&\frac{(-1)^n (k_\rho l_\perp)^{2n+m} e^{-(k_\rho l_\perp)^2/4}}{2^{2n+m+1}\pi\sqrt{n!(n+m)!}},
\end{eqnarray}
with $J_m$ the Bessel function of the first kind. Integrating in Eq.~\eqref{eq:U-1} we obtain for $\epsilon_\text{dd}  =1$ Eq. (2) of the main text.

\section{Derivation of equation (4)}

Equation (4) of the main text is obtained extending the arguments of Hugenholtz and Pines~(HP)~\cite{Hugenholtz1959} to 
the quasi-1D geometry under consideration. We introduce the Green's function, 
$G_{n_r,n'_r,m}(k,t-t')\delta(k-k')\delta_{m,m'}=-i\langle\psi_0 | \hat T \hat a_{n_r,m,k}(t) \hat a_{n'_r,m',k'}^\dag (t') |\psi_0\rangle$, 
with $\hat T$ the time ordering operator, and $\hat a_{n_r,m,k}$ the annihilation operator of a particle in the mode $(n_r,m,k)$.
We apply the HP formalism to obtain the relation:
\begin{eqnarray}
&& \frac{E_0}{L} - \frac{\mu n_\text{1D}  }{2} = \nonumber \\
&&\sum_j \int \!\frac{dk}{2\pi} \!\int_C\!\! \frac{-id\epsilon}{2\pi} \frac{1}{2} \left [ \epsilon\!+\!E_{n_r m}(k) \right ] \! G_{n_r,n_r,m}(k,\epsilon),
\end{eqnarray}
where $n_\text{1D}  $ is the 1D density, $L$ is the quantization length, $E_0$ is the ground-state energy, $\mu$ the chemical potential,  
$G_{n_r,n_r,m}(p,\epsilon)=\int_{-\infty}^{+\infty}G_{n_r,n_r,m}(k,t-t')e^{i\epsilon t}$, $E_{n_r m}(k)=\hbar\omega_\perp (2n_r +m)+\hbar^2k^2/2M$, 
and $\int_C$ denotes integration over the upper half complex plane. This expression, which is the equivalent for our problem of  Eq.~(4.10) of HP, 
is valid at any order in perturbation theory. As in HP, the first order~(LHY) correction 
beyond mean-field is obtained by expressing the Green's function as a function of the Bogoliubov modes~(obtained from Eq.~(3) of the main text). We obtain 
then the equivalent of Eq.~(7.18) of HP applied to our problem: 
\begin{widetext}
\begin{eqnarray}
\frac{\Delta E_\text{LHY}  }{L}-\frac{\Delta\mu_\text{LHY}   n_\text{1D}  }{2} &=& \sum_m \int \frac{dk}{2\pi} \int_C \frac{-id\epsilon}{2\pi} \frac{1}{2} \sum_{n_r} \left (\epsilon\!+\!E_{n_r m}(k)\right ) 
\sum_\nu \left (\frac{1}{\epsilon-\xi_\nu+i\delta} (\vec u_\nu)_{n_r}^2 +\frac{1}{-\epsilon-\xi_\nu+i\delta} (\vec v_\nu)_{n_r}^2 \right ) \nonumber \\
&=& \frac{1}{2}\sum_m \int \frac{dk}{2\pi} \sum_\nu \sum_{n_r }\left [ E_{n_r m}(k) - \xi_\nu \right ] (\vec v_\nu)_{n_r}^2,
\end{eqnarray}
\end{widetext}
which is Eq.~(4) of the main text.

\section{Derivation of equation (6)}

Let $\hat \psi(\vec r)=\psi_0(\vec r)+\hat {\delta\psi}(\vec r)$, with $\psi_0(\vec r)=\langle \hat \psi(\vec r)\rangle$ the condensate wavefunction, and $\hat {\delta\psi}(\vec r)$ the fluctuations. 
Up to first non-vanishing order in the quantum fluctuations, we may express:
\begin{eqnarray}
\langle \hat \psi^\dag(\vec r)^2  \hat \psi (\vec r)^2\rangle &\simeq& n(\vec r)^2 + n(\vec r) W(\vec r), \\ 
\langle \hat \psi^\dag(\vec r)^3  \hat \psi (\vec r)^3\rangle &\simeq& n(\vec r)^3 + 3n(\vec r)^2 W(\vec r), 
\end{eqnarray}
with $n(\vec r)=\langle \hat \psi^\dag(\vec r)  \hat \psi (\vec r)\rangle$ and 
$W(\vec r)=\langle \hat{\delta\psi}(\vec r)^2  + \hat{\delta\psi}^\dag(\vec r)^2  + 2 \hat{\delta\psi}^\dag(\vec r)\hat{\delta\psi}(\vec r) \rangle$
Hence:
\begin{eqnarray}
&&\!\!\!\!\int d^3 r  \, n(\vec r)^2 \left [ \frac{\langle \hat \psi^\dag(\vec r)^3  \hat \psi (\vec r)^3\rangle}{n(\vec r)^3} -1\right ]  \nonumber \\ 
&=&\!\!3\!\int \!\! d^3 r \! \left ( 
\langle \hat \psi^\dag(\vec r)^2  \hat \psi (\vec r)^2\rangle - n(\vec r)^2 \right )\! =\! 6\frac{\partial \Delta E_\text{LHY}  }{\partial g_\text{3D}  },
\label{eq:HF}
\end{eqnarray}
where in the last equality we employed the Hellmann-Feynmann theorem. We may approximate in the top line of Eq.~\eqref{eq:HF} 
$n(\vec r)\simeq |\psi_0(\vec r)|^2=R_{00}(\rho)^2/L$. Using the definition of $g_\text{1D}  $ we then obtain Eq.~(6) of the main text.

%% file: fig2-latex.pdf_tex
\begingroup%
  \makeatletter%
  \providecommand\color[2][]{%
    \errmessage{(Inkscape) Color is used for the text in Inkscape, but the package 'color.sty' is not loaded}%
    \renewcommand\color[2][]{}%
  }%
  \providecommand\transparent[1]{%
    \errmessage{(Inkscape) Transparency is used (non-zero) for the text in Inkscape, but the package 'transparent.sty' is not loaded}%
    \renewcommand\transparent[1]{}%
  }%
  \providecommand\rotatebox[2]{#2}%
  \ifx\svgwidth\undefined%
    \setlength{\unitlength}{536.00001526bp}%
    \ifx\svgscale\undefined%
      \relax%
    \else%
      \setlength{\unitlength}{\unitlength * \real{\svgscale}}%
    \fi%
  \else%
    \setlength{\unitlength}{\svgwidth}%
  \fi%
  \global\let\svgwidth\undefined%
  \global\let\svgscale\undefined%
  \makeatother%
  \begin{picture}(1,0.71583578)%
    \put(0,0){\includegraphics[width=\unitlength,page=1]{fig2-latex.pdf}}%
    \put(0.71263043,0.63453033){\color[rgb]{0,0,0}\rotatebox{59.3928122}{\makebox(0,0)[b]{\smash{$\propto n_\text{1D}^{3/2}$}}}}%
    {\footnotesize\put(0.66016296,0.26727726){\color[rgb]{0,0,0}\makebox(0,0)[rb]{\smash{$-0.2$}}}%
    \put(0.66016296,0.19224724){\color[rgb]{0,0,0}\makebox(0,0)[rb]{\smash{$-0.4$}}}}%
    \put(0.06342049,0.63782356){\color[rgb]{0,0,0}\makebox(0,0)[rb]{\smash{$3$}}}%
    \put(0.06316974,0.5020626){\color[rgb]{0,0,0}\makebox(0,0)[rb]{\smash{$2$}}}%
    \put(0.06356377,0.36669556){\color[rgb]{0,0,0}\makebox(0,0)[rb]{\smash{$1$}}}%
    \put(0.06336676,0.23172263){\color[rgb]{0,0,0}\makebox(0,0)[rb]{\smash{$0$}}}%
    \put(0.0750838,0.0959616){\color[rgb]{0,0,0}\makebox(0,0)[rb]{\smash{$-1$}}}%
    \put(0.68949969,0.03790191){\color[rgb]{0,0,0}\makebox(0,0)[b]{\smash{$0.8$}}}%
    \put(0.53749397,0.03790191){\color[rgb]{0,0,0}\makebox(0,0)[b]{\smash{$0.6$}}}%
    \put(0.38523739,0.03790191){\color[rgb]{0,0,0}\makebox(0,0)[b]{\smash{$0.4$}}}%
    \put(0.23362569,0.03790191){\color[rgb]{0,0,0}\makebox(0,0)[b]{\smash{$0.2$}}}%
    \put(0.08142295,0.03790191){\color[rgb]{0,0,0}\makebox(0,0)[b]{\smash{$0.0$}}}%
    \put(0.4964427,0.00752935){\makebox(0,0)[lb]{\smash{$n_\text{1D}a$}}}%
    \put(0.03041213,0.39411279){\color[rgb]{0,0,0}\rotatebox{90}{\makebox(0,0)[b]{\smash{$\Lambda(n_\text{1D}a)$}}}}%
    \put(0.84055612,0.03790191){\color[rgb]{0,0,0}\makebox(0,0)[b]{\smash{$1.0$}}}%
    {\footnotesize
    \put(0.82655623,0.10382347){\makebox(0,0)[b]{\smash{$n_\text{1D}a$}}}%
    \put(0.56595537,0.25116139){\rotatebox{90}{\makebox(0,0)[b]{\smash{$\Delta E_\text{LHY}/N$}}}}%
    \put(0.21672031,0.66776421){\color[rgb]{0,0,0}\makebox(0,0)[rb]{\smash{$1.0$}}}%
    \put(0.21646956,0.54974277){\color[rgb]{0,0,0}\makebox(0,0)[rb]{\smash{$0.2$}}}%
    \put(0.21666665,0.48544297){\color[rgb]{0,0,0}\makebox(0,0)[rb]{\smash{$-0.2$}}}%
    \put(0.21874271,0.43198296){\color[rgb]{0,0,0}\makebox(0,0)[b]{\smash{$-3.0$}}}%
    \put(0.36230744,0.42984346){\color[rgb]{0,0,0}\makebox(0,0)[b]{\smash{$-2.0$}}}%
    \put(0.49879329,0.43369098){\color[rgb]{0,0,0}\makebox(0,0)[b]{\smash{$-1.0$}}}%
    \put(0.4151925,0.38755341){\color[rgb]{0,0,0}\makebox(0,0)[b]{\smash{$\log_{10}(n_\text{1D}a)$}}}%
    \put(0.12495895,0.57325692){\color[rgb]{0,0,0}\rotatebox{90}{\makebox(0,0)[b]{\smash{$\log_\text{10}(|\Lambda|/n_\text{1D}a)$}}}}%
    \put(0.21646956,0.60937192){\color[rgb]{0,0,0}\makebox(0,0)[rb]{\smash{$0.6$}}}%
    \put(0.66016296,0.34211024){\color[rgb]{0,0,0}\makebox(0,0)[rb]{\smash{$0.0$}}}%
    \put(0.92391858,0.13475929){\color[rgb]{0,0,0}\makebox(0,0)[b]{\smash{$0.6$}}}%
    \put(0.83732951,0.13436524){\color[rgb]{0,0,0}\makebox(0,0)[b]{\smash{$0.4$}}}%
    \put(0.75342849,0.13475929){\color[rgb]{0,0,0}\makebox(0,0)[b]{\smash{$0.2$}}}%
    \put(0.66555164,0.13475929){\color[rgb]{0,0,0}\makebox(0,0)[b]{\smash{$0.0$}}}%
    }
  \end{picture}%
\endgroup%

%% file: fig3-latex.pdf_tex
\begingroup%
  \makeatletter%
  \providecommand\color[2][]{%
    \errmessage{(Inkscape) Color is used for the text in Inkscape, but the package 'color.sty' is not loaded}%
    \renewcommand\color[2][]{}%
  }%
  \providecommand\transparent[1]{%
    \errmessage{(Inkscape) Transparency is used (non-zero) for the text in Inkscape, but the package 'transparent.sty' is not loaded}%
    \renewcommand\transparent[1]{}%
  }%
  \providecommand\rotatebox[2]{#2}%
  \ifx\svgwidth\undefined%
    \setlength{\unitlength}{523.42858887bp}%
    \ifx\svgscale\undefined%
      \relax%
    \else%
      \setlength{\unitlength}{\unitlength * \real{\svgscale}}%
    \fi%
  \else%
    \setlength{\unitlength}{\svgwidth}%
  \fi%
  \global\let\svgwidth\undefined%
  \global\let\svgscale\undefined%
  \makeatother%
  \begin{picture}(1,0.73554774)%
    \put(0,0){\includegraphics[width=\unitlength,page=1]{fig3-latex.pdf}}%
    {\footnotesize
    \put(0.9640703,0.68093398){\color[rgb]{0,0,0}\makebox(0,0)[rb]{\smash{$0.2$}}}%
    \put(0.9640703,0.49687193){\color[rgb]{0,0,0}\makebox(0,0)[rb]{\smash{$-0.2$}}}%
    \put(0.96333306,0.12874803){\color[rgb]{0,0,0}\makebox(0,0)[rb]{\smash{$-1.0$}}}%
    \put(0.96333306,0.31280998){\color[rgb]{0,0,0}\makebox(0,0)[rb]{\smash{$-0.6$}}}%
    \put(0.96379941,0.58890292){\color[rgb]{0,0,0}\makebox(0,0)[rb]{\smash{$0.0$}}}%
    \put(0.96337821,0.4046755){\color[rgb]{0,0,0}\makebox(0,0)[rb]{\smash{$-0.4$}}}%
    \put(0.96382958,0.22077906){\color[rgb]{0,0,0}\makebox(0,0)[rb]{\smash{$-0.8$}}}%
    \put(0.23977703,0.02696065){\color[rgb]{0,0,0}\makebox(0,0)[b]{\smash{$0.98$}}}%
    \put(0.40396041,0.02696065){\color[rgb]{0,0,0}\makebox(0,0)[b]{\smash{$1.02$}}}%
    \put(0.48205193,0.02695278){\color[rgb]{0,0,0}\makebox(0,0)[b]{\smash{$1.04$}}}%
    \put(0.64062079,0.02696065){\color[rgb]{0,0,0}\makebox(0,0)[b]{\smash{$1.08$}}}%
    \put(0.07631237,0.02695278){\color[rgb]{0,0,0}\makebox(0,0)[b]{\smash{$0.94$}}}%
    \put(0.56361872,0.02696065){\color[rgb]{0,0,0}\makebox(0,0)[b]{\smash{$1.06$}}}%
    \put(0.32227469,0.02696065){\color[rgb]{0,0,0}\makebox(0,0)[b]{\smash{$1.00$}}}%
    \put(0.15669583,0.02696065){\color[rgb]{0,0,0}\makebox(0,0)[b]{\smash{$0.96$}}}%
    \put(0.06736313,0.06056064){\color[rgb]{0,0,0}\makebox(0,0)[rb]{\smash{$30$}}}%
    \put(0.06735667,0.12939711){\color[rgb]{0,0,0}\makebox(0,0)[rb]{\smash{$35$}}}%
    \put(0.06782522,0.20828499){\color[rgb]{0,0,0}\makebox(0,0)[rb]{\smash{$40$}}}%
    \put(0.06781877,0.28733835){\color[rgb]{0,0,0}\makebox(0,0)[rb]{\smash{$45$}}}%
    \put(0.06711667,0.36655724){\color[rgb]{0,0,0}\makebox(0,0)[rb]{\smash{$50$}}}%
    \put(0.06711022,0.4456106){\color[rgb]{0,0,0}\makebox(0,0)[rb]{\smash{$55$}}}%
    \put(0.06736313,0.52466398){\color[rgb]{0,0,0}\makebox(0,0)[rb]{\smash{$60$}}}%
    \put(0.06735667,0.60371738){\color[rgb]{0,0,0}\makebox(0,0)[rb]{\smash{$65$}}}%
    \put(0.06693183,0.68262029){\color[rgb]{0,0,0}\makebox(0,0)[rb]{\smash{$70$}}}}%
    \put(0,0){\includegraphics[width=\unitlength,page=2]{fig3-latex.pdf}}%
    \put(0.51597687,0.71433425){\color[rgb]{0,0,0}\makebox(0,0)[b]{\smash{\textsc{Soliton}}}}%
    \put(0.1880193,0.7216467){\color[rgb]{0,0,0}\rotatebox{-0.03630583}{\makebox(0,0)[b]{\smash{\textsc{Attractive LHY}}}}}%
    \put(0,0){\includegraphics[width=\unitlength,page=3]{fig3-latex.pdf}}%
    \put(0,0){\includegraphics[width=\unitlength,page=4]{fig3-latex.pdf}}%
    \put(0.19609945,0.34776414){\color[rgb]{1,1,1}\makebox(0,0)[b]{\smash{\footnotesize\textsc{1D Droplet}}}}%
    \put(0,0){\includegraphics[width=\unitlength,page=5]{fig3-latex.pdf}}%
    \put(0.52453899,0.34788245){\color[rgb]{0,0,0}\makebox(0,0)[b]{\smash{\footnotesize\textsc{Repulsive LHY}}}}%
    \put(1.00275737,0.39159464){\color[rgb]{0,0,0}\rotatebox{90}{\makebox(0,0)[b]{\smash{$\log_{10}(n_\text{1D}a)$}}}}%
    \put(0.38790999,-0.00170874){\color[rgb]{0,0,0}\makebox(0,0)[b]{\smash{$\varepsilon_\text{dd}$}}}%
    \put(0.01521004,0.37464695){\color[rgb]{0,0,0}\rotatebox{90}{\makebox(0,0)[b]{\smash{$l_\perp/a$}}}}%
    \put(0.8048916,0.38001734){\color[rgb]{0,0,0}\rotatebox{90}{\makebox(0,0)[b]{\smash{\textsc{3D Droplet}}}}}%
  \end{picture}%
\endgroup%

%% file: LHY1D.bbl
\begin{thebibliography}{99}
\bibitem{footnote-alpha} For a general angle $\alpha$ between the dipole moment and the $z$ axis, 
$\varepsilon_\text{dd}  =\mu_0\mu_D^2 (3\cos^2\alpha-1)/{6 g_\text{3D}}$. Hence, contrary to 3D systems, the crucial parameter $\varepsilon_\text{dd}  $,
may be externally reduced in 1D by modifying $\alpha$. This may be particularly useful in the case of dysprosium, where $\varepsilon_\text{dd}>1$ for $\alpha=0$, 
for which the crossover through $\varepsilon_\text{dd}=1$ may be realized without the need of a Feshbach resonance.
\bibitem{footnote-1D} In an homogeneous quasi-1D dipolar condensate, $\mu=g_\text{1D}n_\text{1D}(1-\varepsilon_\text{dd})$. Hence 
$\mu/{\hbar\omega_\perp} = 2 n_\text{1D} a (1-\varepsilon_\text{dd})$.
\bibitem{footnote-comment-edd} In the vicinity of $\varepsilon_\text{dd}  =1$, which is our regime of interest,  quantum corrections variate only very slightly with $\varepsilon_\text{dd}  $.
\bibitem{footnote-g3} Note that the three-body losses are proportional to $g^{(3)}n(\vec r)^2$. Hence, the proper transversal average is performed by averaging over $\psi(x,y)^4$.
\end{thebibliography}
